\documentclass[journal]{IEEEtran}
\usepackage{ifpdf}

\usepackage{cite}

\ifCLASSINFOpdf
\usepackage[pdftex]{graphicx}
\graphicspath{{../Figures/}}
\DeclareGraphicsExtensions{.pdf}
\else
\fi

\ifCLASSOPTIONcompsoc
\usepackage[caption=false,font=normalsize,labelfont=sf,textfont=sf]{subfig}
\else
\usepackage[caption=false,font=footnotesize]{subfig}
\fi

\usepackage{multirow}

\usepackage{amssymb}
\usepackage{amsfonts}
\usepackage{amsmath}
\usepackage{algorithm}
\usepackage{algorithmic}
\usepackage{amsthm}

\usepackage{easyReview}

\usepackage{array}

\usepackage{url}

\makeatletter
\let\NAT@parse\undefined
\makeatother
\usepackage{hyperref}  
\usepackage{color}

\begin{document}
	
	\title{Prototyping and Real-world Field Trials of RIS-aided Wireless Communications}
	
	\author{Xilong Pei, Haifan Yin,~\IEEEmembership{Member,~IEEE}, Li Tan, Lin Cao, and Taorui Yang
		\thanks{X. Pei, H. Yin, L. Tan, L. Cao and T. Yang are with the School of Electronic Information and Communications, Huazhong University of Science and Technology, Wuhan, China. E-mail: \{pei, yin, ltan, lincao, try\}@hust.edu.cn.}
		\thanks{The corresponding author is Haifan Yin.}
		\thanks{This work was supported in part by the National Key Research and Development Program of China under Grant 2020YFB1806904, in part by by the National Natural Science Foundation of China under Grants 62071191, 62071192 and 1214110.}
	}
	
	
	\maketitle
	
	\begin{abstract}
		Reconfigurable intelligent surface (RIS) is a promising technology that has the potential to change the way we interact with the wireless propagating environment. In this paper, we design and fabricate an RIS system that can be used in the fifth generation (5G) mobile communication networks. We also propose a practical two-step spatial-oversampling codebook algorithm for the beamforming of RIS,  which is based on the spatial structure of the wireless channel. This algorithm has much lower complexity compared to the two-dimensional full-space searching-based codebook, yet with only negligible performance loss. Then, a series of experiments are conducted with the fabricated RIS systems, covering the office, corridor, and outdoor environments, in order to verified the effectiveness of RIS in both laboratory and current 5G commercial networks. In the office and corridor scenarios, the 5.8 GHz RIS provided a 10-20 dB power gain at the receiver. In the outdoor test, over 35 dB power gain was observed with RIS compared to the non-deployment case. However, in commercial 5G networks, the 2.6 GHz RIS improved indoor signal strength by only 4-7 dB. The experimental results indicate that RIS achieves higher power gain when transceivers are equipped with directional antennas instead of omni-directional antennas.
	\end{abstract}
	
	
	\IEEEpeerreviewmaketitle
	
	\section{Introduction}
	
	\IEEEPARstart{I}{n} the past few years, the fifth generation (5G) mobile communication technology has been widely deployed in countries around the world.
	To improve communications performance, 5G networks employ a variety of technologies, such as massive multiple-input multiple-output (MIMO), millimeter wave (mmWave) wireless communications and ultra-dense network (UDN) \cite{marzetta2010noncooperative, lu2014overview, tse2005fundamentals, ahmed2018survey, rappaport2013millimeter}.
	While these technologies have improved spectrum efficiency, they have also increased costs and power consumption.
	For example, in massive MIMO, base stations (BSs) enhance network capacity by multiplying high-cost radio-frequency (RF) chains. 
	Recently, the sixth generation (6G) mobile communication technology has attracted much attention from both academia and industry.
	6G networks are expected to enable new applications and services beyond current mobile use scenarios, such as virtual and augmented reality (VR/AR), holographic telepresence \cite{giordani2020toward}.
	However, some challenges need to be solved, e.g., how to maximize the coverage and improve the spectrum efficiency with lower deployment costs.
	
	Reconfigurable intelligent surfaces (RISs) have emerged as a prospective technology that can enhance signal coverage and improve link capacity at a low cost \cite{liu2021reconfigurable, huang2019reconfigurable}.
	Therefore, RIS-aided communication is a candidate for future 6G systems \cite{basar2020reconfigurable}.
	RIS is a two-dimensional array plane based on a sub-wavelength structure \cite{cui2009ruopeng}, that can achieve flexible and effective control of the propagation direction, polarization direction, and propagation mode of electromagnetic waves \cite{gao2018reconfigurable, tao2017reconfigurable}.
	RIS-aided communication solutions offer the advantages of low production costs, lightweight, and low power consumption, and can be easily deployed both indoors and outdoors (e.g., on ceilings or building facades) \cite{liaskos2018new, wu2021intelligent}.
	Furthermore, it is possible to integrate RIS with current communication systems in a nearly transparent way, i.e., without requiring the end-user equipment to undergo any complicated hardware or software updates \cite{pei2021ris}.
	
	There have been numerous theoretical and practical studies on RIS in recent years, including modeling, channel state information (CSI) estimation, hardware implementations, system prototyping, and experimental results \cite{wu2019intelligent, tang2020wireless, abeywickrama2020intelligent}.
	Based on the existing theoretical research results, many research teams developed RIS prototypes.
	In 2014, programmable metamaterials were introduced and implemented, which is the basic hardware architecture of RIS \cite{cui2014coding}.
	In \cite{arun2019rfocus}, researchers developed the RFocus lens, which is a fully passive prototype system that has the ability of smart beamforming for the incident signals.
	According to their indoor measurements, RFocus can enhance signal strength by an average of 9.5 times and double the channel capacity.
	Since 2020, some academic and research organizations have developed RIS-aided communication systems and performed experiments in the laboratory and in the real world.
	The authors of \cite{dai2020reconfigurable} proposed an RIS prototype with 256 2-bit elements. At 2.3 GHz, it achieved a 21.7 dBi antenna gain in the anechoic chamber.
    In \cite{araghi2022reconfigurable}, a measurement campaign was carried out to simulate a blocking situation in the actual world.
	The results showed that the RIS enhanced the received signal strength by over 15 dB compared to the power-off case.
    Few studies have examined the performance of RIS in commercial networks to date.
	In \cite{sang2022coverage, liu2022simulation}, the authors conducted field trials on RIS performance of enhancing coverage in current 5G commercial networks.
	Experimental results prove the effectiveness of RISs in resolving coverage issues and enhancing signal quality.
	These trials have contributed to the theoretical researches and commercial applications of RIS.
   However, the test scenarios in these studies are not sufficiently comprehensive, and there is a lack of comparison between practical beamforming algorithms.
	
	In this paper, we present two RIS prototype systems that can perform real-time beamforming.
	One targets the 5.8 GHz band to enhance wireless local area network (WLAN) signals, and the other targets the 2.6 GHz band for the China Mobile 5G Network.
	We constructed the verification systems and conducted an extensive series of tests in various environments, including microwave anechoic chambers, indoor scenarios, outdoor scenarios, and commercial network environments.
	All field trials were carried out under realistic conditions (e.g., wind blowing, tree branches swaying, pedestrians walking, etc.).
	In outdoor environments, the power gain at the receiver could exceed 35 dB compared to no-deployment case.
	With the 2.6 GHz RIS, we also discovered a 4 to 7 dB gain in the 5G commercial network.
	
	The main contributions of our work are as follows:
	First, we propose a low-complexity two-step spatial-oversampling codebook algorithm, which is based on the spatial structure of the channel.
	Experimental results show that the effect of the two-step algorithm is similar to that of the two-dimensional angle-based codebook, which has much higher complexity.
	Second, we assess the performance of various algorithms and analyze their strengths and weaknesses.
	Third, we examine the performance gap between directional and omnidirectional antennas in the RIS-aided wireless communication systems.
	We find that the gain of RIS is inadequate when omnidirectional antennas are equipped at the transceivers, which poses a practical challenge for RIS in the future commercialization.

	The rest of this paper is organized as follows.
	Section \ref{sec:prototypes} describes the composition and implementation of the RIS systems.
	A codebook design algorithm is also presented in this section.
	The experiments based on the 5.8 GHz prototype system is demonstrated in the following section, along with discussions and analyses of the test results for the aforementioned scenarios.
	In Section \ref{sec:2.6GHz_RIS_test}, we test the 2.6 GHz prototype system in the lab and in the commercial network.
	Finally, the conclusions are drawn in Section \ref{sec:conclusion}.
	
	\section{RIS Prototype Systems}\label{sec:prototypes}
	In this section, we give a brief introduction to the 5.8 GHz RIS system and detail our 2.6 GHz RIS hardware prototype.
	
	\begin{table}[t!]
		\renewcommand{\arraystretch}{1.3}
		\caption{5.8 GHz RIS Specifications Table}
		\label{tab:5.8GHz_RIS_Specifications}
		\centering
		\begin{tabular}{|c|c|}
			\hline\hline
			\textbf{Parameter}       &                   \textbf{Value}                   \\ \hline
			Type              &                 Reflective surface                 \\ \hline
			Center frequency        &                      5.8 GHz                       \\ \hline
			Operating frequency range   &   $ [5.7 \,\mathrm{GHz} , 5.9 \,\mathrm{GHz}] $    \\ \hline
			Aperture         & $ 78.65 \,\mathrm{cm} \times 20.54 \,\mathrm{cm} $ \\ \hline
			Polarization direction     &              Horizontal               \\ \hline
			Number of RIS units      &                    $ L = 1100 $                    \\ \hline
			Number of rows         &                     $ N = 20 $                     \\ \hline
			Number of columns       &                     $ M = 55 $                     \\ \hline
			Switching rate         &                       1 MHz                        \\ \hline
			Beam scanning range      &                   $ 160^\circ $                    \\ \hline
			$ 3 \,\mathrm{dB} $ beamwidth &                   $ 5.2^\circ $                    \\ \hline\hline
		\end{tabular}
	\end{table}
	
	We have previously demonstrated the 5.8 GHz RIS prototype system in \cite{pei2021ris}.
	Here, we provide a brief overview of the system, which consists of a programmable metasurface and its control circuit as the RIS part, and two host computers and two universal software radio peripherals (USRPs) as the transceiver part.
	
	The specifications of the 5.8 GHz RIS can be found in Table \ref{tab:5.8GHz_RIS_Specifications}.
	The RIS panel consists of $55\times20$ units arranged periodically.
	It operates at a central frequency of 5.8 GHz with a bandwidth of approximately 200 MHz.
	By applying an external bias voltage to each structural unit, the junction capacitance of the varactor can be adjusted, which allows dynamic control of the reflection coefficient.
	The RIS can achieve a beam angle adjustment range up to 160 degrees at 5.8 GHz.
	For the transceiver, the host computer converts the video data into a bitstream, modulates it using orthogonal frequency division multiplexing (OFDM), and transmits it through USRP at the transmitting end.
	At the receiving end, the host collects signals via a horn antenna and performs two tasks: decoding the signals into a bitstream and estimating the channel for beamforming.
	The signal power is calculated using the pilot and the power information is fed back to the RIS controller.
	
	Next, we will introduce the detailed design of the 2.6 GHz RIS system.
	
	\subsection{RIS Board}
\begin{figure}[t!]
		\centering
		\includegraphics[width=\linewidth]{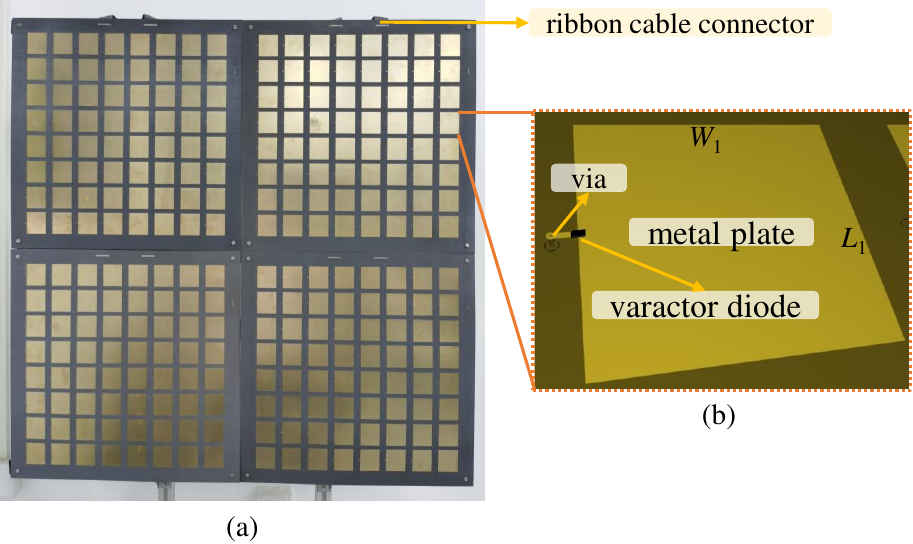}
		\caption{Fabricated 2.6 GHz RIS: (a) front view; (b) unit cell.}
		\label{fig:2.6GRIS}
	\end{figure}

	We designed an $8 \times 8$ RIS panel for 2.6 GHz band with specifications shown in Table \ref{tab:2.6GHz_RIS_Specifications}.
	Fig. \ref{fig:2.6GRIS} (a) show its front view (four pieces of RIS are assembled together in the figure).
	Fig. \ref{fig:2.6GRIS} (b) illustrates the structure of the RIS unit, which comprises a rectangular patch and a varactor diode on the top layer, a ground plane in the middle layer, and a control line on the bottom layer.
	The geometric parameters of the unit cell are $ L_1=42 $ mm and $ W_1=37 $ mm.
	The substrate of the patches is made of F4B material with a thickness of 2 mm, while the DC part behind the ground plane uses FR4 material with a thickness of 0.2 mm.
	A through hole with a diameter of 0.25 mm connects one end of the diode to the control line.
	We chose varactor diodes SMV1408-040LF by Skyworks Solutions, which can be modeled as a series of Resistor-Inductor-Capacitor (RLC) elements.
	When the bias voltage changes, the varactor diode achieves different capacitance values, thereby changing the circuit characteristics.
	
\begin{figure}[t!]
		\centering
		\includegraphics[width=\linewidth]{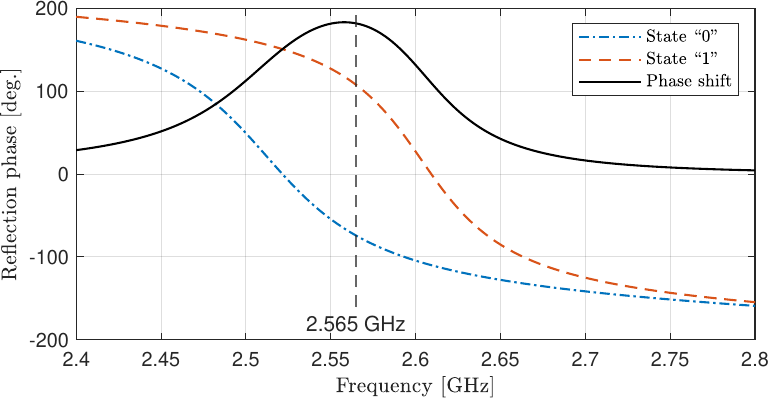}
		\caption{The simulated reflection phase of the unit cell in different states, where the phase shift between the ``0'' state and ``1'' state is 180 degrees at 2.565 GHz.}
		\label{fig:2.6G_Sim}
	\end{figure}

	The characteristics of the proposed 2.6 GHz RIS unit cell were simulated using the commercial electromagnetic solver CST Microwave Studio 2019.
	Considering the practical application scenarios, the incidence angle is $0^\circ$ (i.e. vertical incidence).
	Under these conditions, we adjusted the state of the diodes to observe the changes in the reflection coefficient.
	The simulated results of the reflection phase and amplitude of the unit cell are presented in Fig. \ref{fig:2.6G_Sim}.
	The phase differences between the two states are 180 degrees at 2.565 GHz, which meets our design requirements.
	
	Based on the unit structure mentioned above, considering the dimensions of the circuit board and the challenges involved in fabrication, the 2.6 GHz RIS was designed with an $8\times8$ unit cell configuration, spanning 45 cm $\times$ 45 cm in size.
	In practical applications, the combination of RIS can be customized to meet specific demands.
	Additionally, a half-wavelength gap is maintained along the edges of the panels, ensuring that the units are positioned at integer multiples of the half wavelength, which simplifies the design of the codebook.
	
	\begin{table}[t!]
		\renewcommand{\arraystretch}{1.3}
		\caption{2.6 GHz RIS Specifications Table}
		\label{tab:2.6GHz_RIS_Specifications}
		\centering
		\begin{tabular}{|c|c|}
			\hline\hline
			\textbf{Parameter}     &                 \textbf{Value}                  \\ \hline
			Type            &               Reflective surface                \\ \hline
			Center frequency      &                     2.6 GHz                     \\ \hline
			Operating frequency range & $ [2.55 \,\mathrm{GHz} , 2.65 \,\mathrm{GHz}] $ \\ \hline
			Size            &  $ 45 \,\mathrm{cm} \times 45 \,\mathrm{cm} $   \\ \hline
			Polarization direction   &             Horizontal             \\ \hline
			Number of RIS units    &                   $ L = 64 $                    \\ \hline
			Number of rows       &                    $ N = 8 $                    \\ \hline
			Number of columns     &                    $ M = 8 $                    \\ \hline
			Switching rate       &                      1 MHz                      \\ \hline\hline
		\end{tabular}
	\end{table}
	
	\subsection{RIS Codebook}
	
	\begin{figure}[t!]
		\centering
		\includegraphics[width=0.7\linewidth]{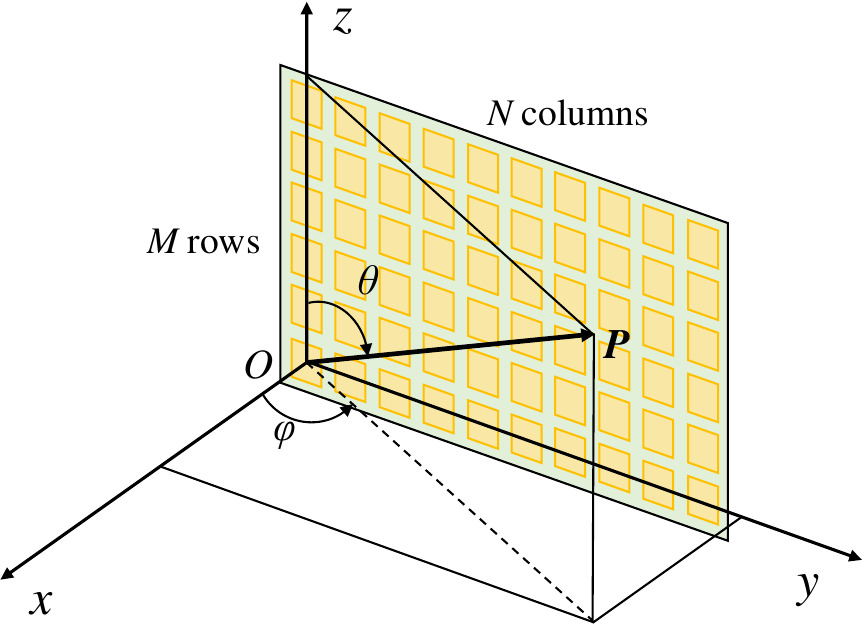}
		\caption{The UPA model of the considered RIS.}
		\label{fig:UPA_model}
	\end{figure}
	
	In \cite{pei2021ris}, a beamforming algorithm based on the Discrete Fourier Transform (DFT) codebook is proposed, which obtains the codeword with the maximum receiving power by searching the codeword in the codebook.
	However, this algorithm cannot precisely adjust the angle of the main lobe in each codeword.
	Therefore, we utilize the array response vector to build the codebook, as it enables a finer adjustment of the reflected beam.
	
	The UPA model of RIS is depicted in Fig. \ref{fig:UPA_model}.
	The RIS is made up of $L=MN$ elements, where $M$ represents the number of RIS elements in the \textit{z}-axis direction, and $N$ represents the number of RIS elements in the \textit{y}-axis direction.
	The first reflecting unit is located at the origin.
	Observe from the origin, the zenith angle and azimuth angle to the transmitter are $\theta$ and $\varphi$, respectively.
	The array response vector $ \mathbf{a}(\theta, \varphi) \in \mathbb{C}^{L \times 1} $ for the reflected wave can be expressed as\cite{3gpp:38.901,yin2020JSAC}:
	\begin{equation}
		\mathbf{a}(\theta, \varphi)=\mathbf{a}_{y}(\theta, \varphi) \otimes \mathbf{a}_{z}(\theta),
		\label{eq:SteeringVector}
	\end{equation}
	where $ \otimes $ denotes the Kronecker product.
	$ \mathbf{a}_{y}(\theta, \varphi) \in \mathbb{C}^{N \times 1}$ and $ \mathbf{a}_{z}(\theta) \in \mathbb{C}^{M \times 1}$ represent the array response vectors of a uniform linear array along the \textit{y}-axis and the \textit{z}-axis respectively.
	As a result, the $i$-th entry of $\mathbf{a}(\theta, \varphi)$ is as follows:
	\begin{equation}
		[\mathbf{a}(\theta, \varphi)]_i=e^{j\phi_{m,n}}=e^{-j\frac{2 \pi}{\lambda}(m d_{z} \cos \theta + n d_{y} \sin \theta \sin \varphi)},
		\label{eq:SteeringVector_i}
	\end{equation}
	where $i=mN+n$ represents the element in $m$-th row and $n$-th column of the RIS.
	
	According to Eq. (\ref{eq:SteeringVector_i}), by changing the zenith angle $\theta$ and the azimuth angle $\varphi$, we generate array response vectors in different directions.
	The spacing and range between angles can be adjusted according to the requirements.
	This allows us to create a more effective codebook after choosing a series of specific angles.
	For example, we choose $p$ azimuth angles $\varphi_1,\varphi_2,\cdots,\varphi_p$ and $q$ zenith angles $\theta_1,\theta_2,\cdots,\theta_q$.
	The generated codebook $\mathbf{F} \in \mathbb{C}^{L \times (pq)}$ is represented as
	\begin{equation}
		\mathbf{F} = [\mathbf{a}(\theta_1, \varphi_1), \mathbf{a}(\theta_1, \varphi_2), \cdots, \mathbf{a}(\theta_q, \varphi_p)].
		\label{eq:codebook}
	\end{equation}
	
	The above method has a higher level of complexity, since two-dimensional search is needed.
	To reduce time complexity, we propose a two-step spatial-oversampling codebook algorithm.
	Firstly, we ignore the phase difference of each element on the \textit{y}-axis,
	that is, the state of each column of reflective elements is consistent with the first column.
	We create the codebook $\mathbf{F}_z \in \mathbb{C}^{M \times q}$ for the first column (\textit{z}-axis) as
	\begin{equation}
		\mathbf{F}_z = [\mathbf{a}_z(\theta_1), \mathbf{a}_z(\theta_2), \cdots, \mathbf{a}_z(\theta_q)],
		\label{eq:Fz}
	\end{equation}
	where $\mathbf{a}_{z}(\theta) =
	\left[1,
	e^{\frac{-j 2 \pi}{\lambda} d_{z} \cdot\cos \theta},
	\cdots,
	e^{\frac{-j 2 \pi}{\lambda} d_{z} \cdot(M-1) \, \cos \theta}
	\right]^\mathrm{T}.$
	
	After $q$ times of scanning, the codeword $\mathbf{a}_z(\theta_{q_{m}})$ with the highest receiving power is selected.
	Similarly, the codebook $\mathbf{F}_y \in \mathbb{C}^{N \times p}$ for the first row (\textit{y}-axis) can be written as
	\begin{equation}
		\mathbf{F}_y = \left[\mathbf{a}_{y}(\theta_{q_{m}}, \varphi_1), \mathbf{a}_{y}(\theta_{q_{m}}, \varphi_2), \cdots, \mathbf{a}_{y}(\theta_{q_{m}}, \varphi_p)\right]
		\label{eq:Fy}
	\end{equation}
	where
	\begin{small}
		\begin{equation*}
			\mathbf{a}_{y}(\theta, \varphi) =  
			\left[1, e^{\frac{-j 2 \pi}{\lambda} d_{y} \, \sin \theta \sin \varphi },
			\cdots,
			e^{\frac{-j 2 \pi}{\lambda} d_{y} \,(N-1) \, \sin \theta \sin \varphi}
			\right]^\mathrm{T}.
		\end{equation*}
	\end{small}
	
	As shown in (\ref{eq:SteeringVector}), the codebook $\mathbf{\hat{F}} \in \mathbb{C}^{L \times p}$ of the second scan can be obtained by calculating the Kronecker product of $\mathbf{F}_y$ and $\mathbf{a}_z(\theta_{q_{m}})$:
	\begin{align}\label{eq:F}
		\mathbf{\hat{F}} &= \mathbf{F}_y \otimes \mathbf{a}_z(\theta_{q_{m}}) \\ 
		&= [\mathbf{a}_{y}(\theta_{q_{m}}, \varphi_1) \otimes \mathbf{a}_z(\theta_{q_{m}}),
		\mathbf{a}_{y}(\theta_{q_{m}}, \varphi_2) \otimes \mathbf{a}_z(\theta_{q_{m}}),
		\cdots,\notag\\
		&\quad\mathbf{a}_{y}(\theta_{q_{m}}, \varphi_p) \otimes \mathbf{a}_z(\theta_{q_{m}})].
	\end{align}
	
	\begin{algorithm}[t!]
		\caption{Two-step Spatial Oversampling Codebook Algorithm}
		\label{alg:twoStepAlgorithm}
		\begin{algorithmic}[1]
			\REQUIRE $p$ azimuth angles $\varphi_1,\varphi_2,\cdots,\varphi_p$, $q$ zenith angles $\theta_1,\theta_2,\cdots,\theta_q$, the feedback of RX signal power $\mathcal{P}_t$.
			\ENSURE The reflection coefficients matrix $\mathbf{R}$.
			\STATE Initialize a reflection coefficients matrix $ \mathbf{R}_0\in \mathbb{C}^{M \times N }$;
			\STATE Receive initial feedback of the RX signal power $ \mathcal{P}_m $;
			\STATE Compute $\mathbf{F}_z$ using Eq. \eqref{eq:Fz};
			\STATE // Horizontal search;
			\FOR {each $i \in [1, q]$}
			\STATE  $ \mathbf{R}_{i} \leftarrow [\overbrace{\mathbf{a}_{z}(\theta_{i}), \mathbf{a}_{z}(\theta_{i}), \cdots, \mathbf{a}_{z}(\theta_{i})}^{N}] $;
			\STATE Receive feedback $ \mathcal{P}_{i} $ under configuration $ \mathbf{R}_{i}$;
			\IF {$ \mathcal{P}_i \ge \mathcal{P}_m$}
			\STATE $ q_m \leftarrow i$; $\mathcal{P}_m \leftarrow \mathcal{P}_i$;
			\ENDIF
			\ENDFOR
			\RETURN{zenith angle $q_m$, highest receiving power $\mathcal{P}_m$;}
			
			\STATE // Vertical search;
			\STATE Compute $\mathbf{F}_y$ using Eq. (\ref{eq:Fy});
			\STATE Compute $\mathbf{\hat{F}}$ using Eq. (\ref{eq:F});
			\STATE Denote $ \mathbf{R} = [\mathbf{r}_{1},\mathbf{r}_{2},\cdots,\mathbf{r}_{N}]$;
			\FOR {each $j \in [1, p]$}
			\STATE $[\mathbf{r}_{1}^\mathrm{T},\mathbf{r}_{2}^\mathrm{T},\cdots,\mathbf{r}_{N}^\mathrm{T}]^\mathrm{T} \leftarrow \mathbf{a}_{y}(\theta_{q_{m}}, \varphi_j) \otimes \mathbf{a}_z(\theta_{q_{m}})$;
			\STATE $ \mathbf{R}_j  \leftarrow [\mathbf{r}_{1},\mathbf{r}_{2},\cdots,\mathbf{r}_{N}]$;
			\STATE Receive feedback $ \mathcal{P}_{j} $ under configuration $ \mathbf{R}_{j}$;
			\IF {$ \mathcal{P}_j \ge \mathcal{P}_m $}
			\STATE $ p_m \leftarrow j$; $\mathcal{P}_m \leftarrow \mathcal{P}_j$; $\mathbf{R}_m \leftarrow \mathbf{R}_j$;
			\ENDIF
			\ENDFOR
			\RETURN{reflection coefficient matrix $ \mathbf{R}_m$.}
		\end{algorithmic}
	\end{algorithm}
	
	The detailed procedures is summarized in Algorithm \ref{alg:twoStepAlgorithm}.
	The proposed approach reduces the original $pq$-times search to $p+q$-times, thus lead to a significant reduction of the complexity.
	
	\subsection{RIS Test Platform}
	
	\begin{figure}[t!]
		\centering
		\includegraphics[width=\linewidth]{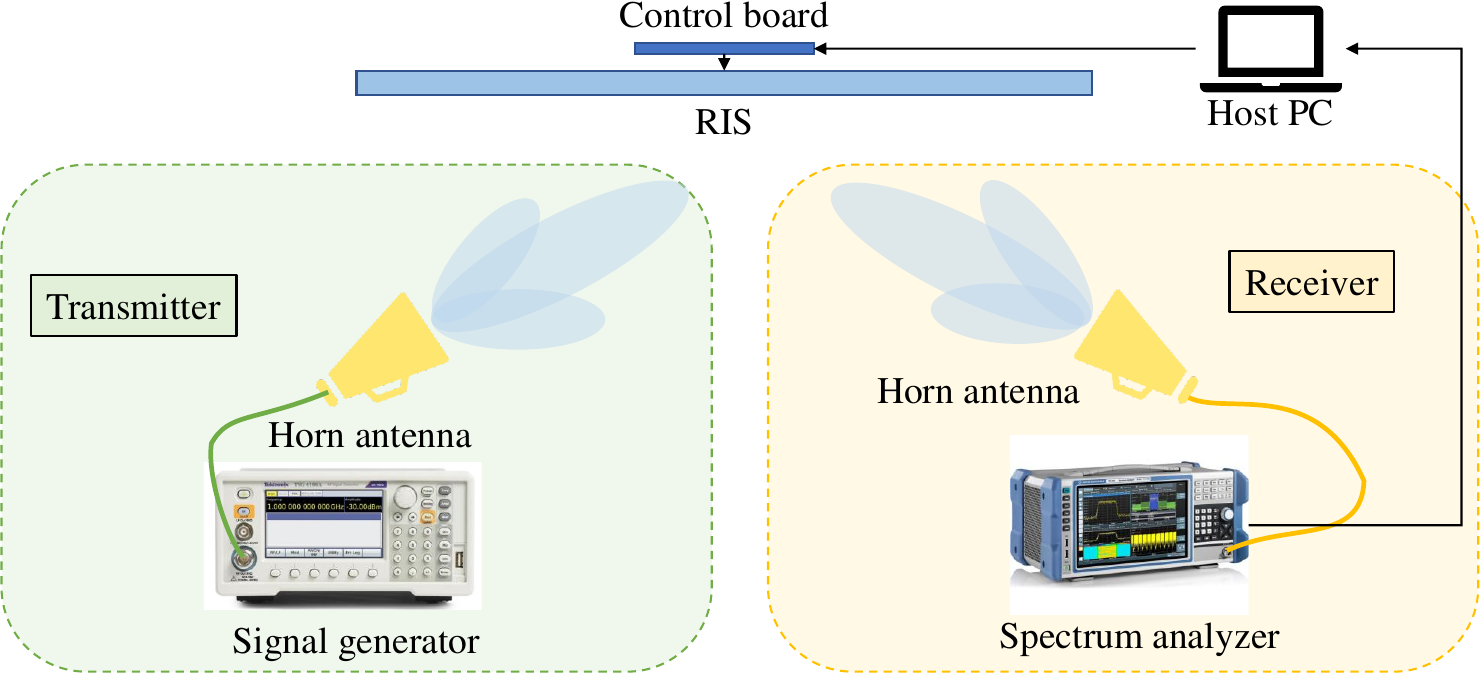}
		\caption{Architecture diagram of the test platform. The host PC obtains power values from the spectrum analyzer through the Ethernet interface and then provides feedback to the RIS control board.}
		\label{fig:test-platform}
	\end{figure}
	
	The block diagram of the test platform is shown in Fig. \ref{fig:test-platform}.
	To increase the measurement accuracy, we use instruments instead of USRPs to generate and receive signals.
	The test platform consists of an RF signal source (Tektronix, TSG4106A), a spectrum analyzer (Rohde \& Schwarz, FPL1007), an RIS panel, an RIS controller, and a host PC.
	The RF signal source emits a sinusoidal RF signal (or a modulated signal with a certain bandwidth), and the spectrum analyzer measures the power (or the total power over a certain bandwidth) of the received signal.
	The host PC reads the power level from the spectrum analyzer and sends it to the RIS controller. The RIS controller runs the beamforming algorithm and optimizes it in real time based on the feedback of the power level. In all experiments, we use standard horn antennas with a 14.8 dBi antenna gain as transmitting and receiving antennas.	
	
	\section{5.8 GHz RIS Field Trials}\label{sec:5.8GHz_RIS_test}
	In this section, we conducted several experiments using our 5.8 GHz RIS prototype system.
	The radiation pattern of the RIS was measured under various angle configurations.
	Subsequently, three major experiments were conducted in a typical office environment, a corridor environment, and an outdoor environment, respectively.
	
	\begin{table}[t!]
		\renewcommand{\arraystretch}{1.3}
		\caption{5.8 GHz RIS Testbed Setup}
		\label{tab:5.8GHz_Testbed_Setup}
		\centering
		\begin{tabular}{|c|c|}
			\hline\hline
			       \textbf{Name}        &              \textbf{Description}              \\ \hline
			RF Vector Signal Generator  &  Tektronix$^\circledR$ TSG4106A (up to 6 GHz)  \\ \hline
			Frequency Spectrum Analyzer &      Rohde \& Schwarz$^\circledR$ FPL1007      \\ \hline
			            VNA             &       Rohde \& Schwarz$^\circledR$ ZNB8        \\ \hline
			       Antenna Gain         &               17.1 dBi @ 5.8 GHz               \\ \hline
			     Antenna Aperture       & $ 169 \,\mathrm{mm} \times 119 \,\mathrm{mm} $ \\ \hline\hline
		\end{tabular}
	\end{table}
	
	\subsection{Radiation Pattern Test}
	
	\begin{figure}[t!]
		\centering
		\includegraphics[width=\linewidth]{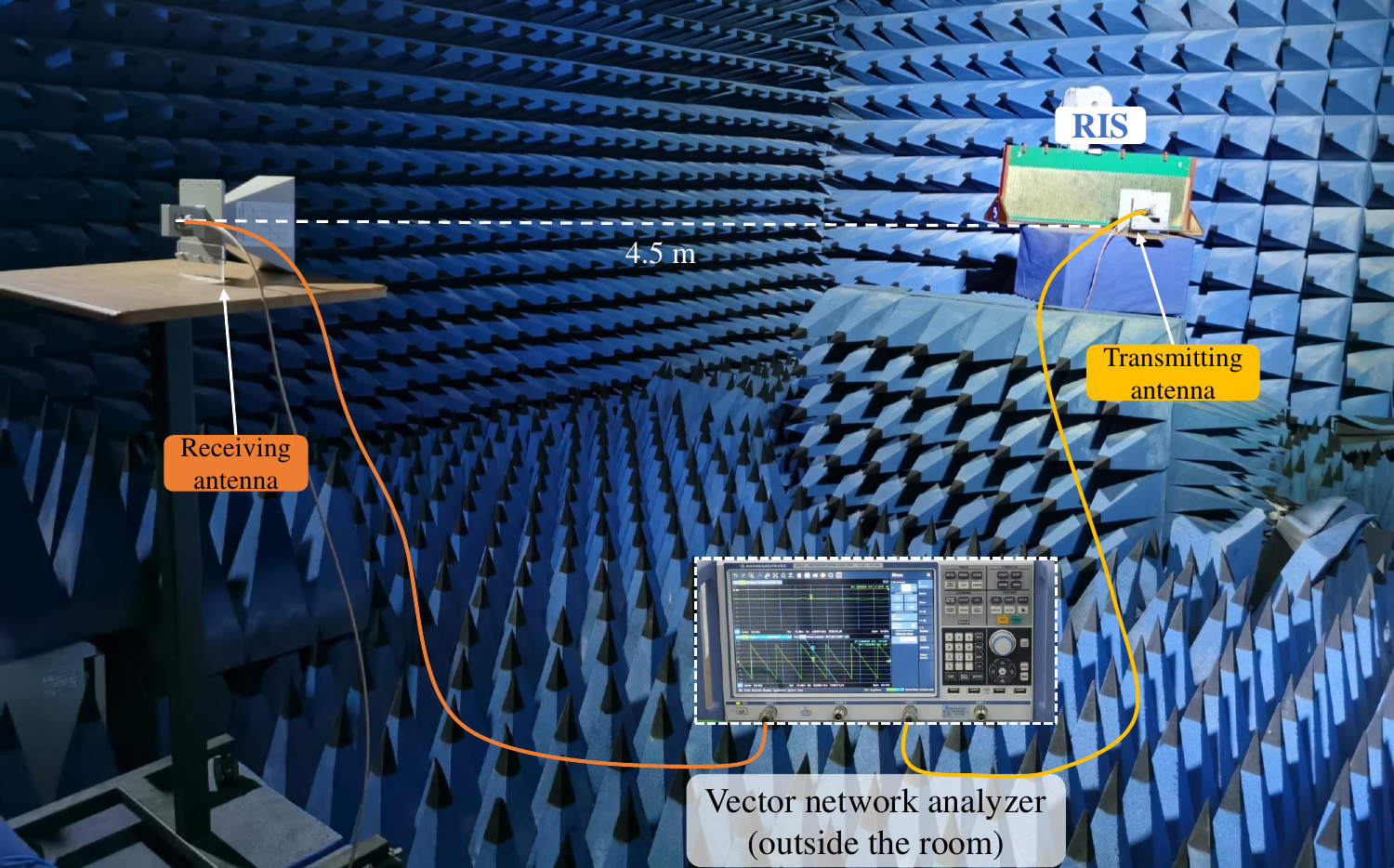}
		\caption{Test scene for the radiation pattern. The experiment is conducted in a microwave anechoic chamber.}
		\label{fig:chamber}
	\end{figure}
	
	The fabricated RIS was tested in a microwave anechoic chamber.
	The chamber dimensions are $ 7.5 \,\mathrm{m} \times 5.5 \,\mathrm{m} \times 3.5 \,\mathrm{m} $.
	As illustrated in Fig. \ref{fig:chamber}, the main testing equipment was a vector network analyzer (VNA) and a turntable system for antenna testing.
	Two horn antennas operating in the 5.7 to 5.9 GHz band were connected to the VNA for transmitting and receiving RF signals respectively.
	The transmitting antenna was fixed on the extension arm of the turntable, which faces the center of the RIS and follows the turntable with the RIS.
	They were separated by 50 centimeters.
	The receiving antenna was mounted on a height-adjustable table to ensure that the two antennas and the center of the RIS were at the same height.
	The distance between the RIS and the receiving antenna was 4.5 m.
	A computer controlled the movement of the turntable and recorded the S parameter from the VNA through the General Purpose Interface Bus (GPIB) protocol during the measurement.
	
\begin{figure}[t!]
		\centering
		\includegraphics[width=\linewidth]{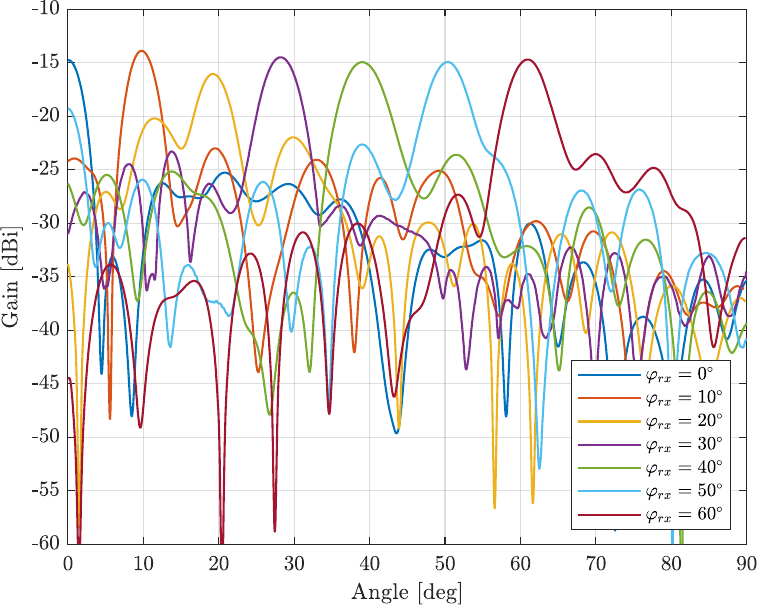}
		\caption{The measured radiation pattern of the RIS with pre-defined reflection coefficients. The incident angle is $0^\circ$ in the measurement, and the desired reflection angle $\varphi_{rx}$ is set as $0^\circ, 10^\circ, 20^\circ, 30^\circ, 40^\circ, 50^\circ, 60^\circ$, respectively.}
		\label{fig:radiation_pattern}
	\end{figure}

	We generate the codebook according to the angle-based codebook with incidence angle of $0^\circ$ and reflection angle of $0^\circ, 10^\circ, 20^\circ, 30^\circ, 40^\circ, 50^\circ, 60^\circ$, respectively.
	Fig. \ref{fig:radiation_pattern} shows the measured radiation patterns, which have peak angles of the main lobes consistent with the desired directions.
	The highest point of the RIS radiation pattern is flat at these angles, and the main lobe gain fluctuates by only 2 dB.
	Moreover, the side lobe is 9.8 dB lower than the main lobe on average.
	These results indicate that the proposed angle-based codebook method is effective in achieving high tuning performance and good stability.
	
	\subsection{Office environment}
	
	\begin{figure}[t!]
		\centering
		\includegraphics[width=\linewidth]{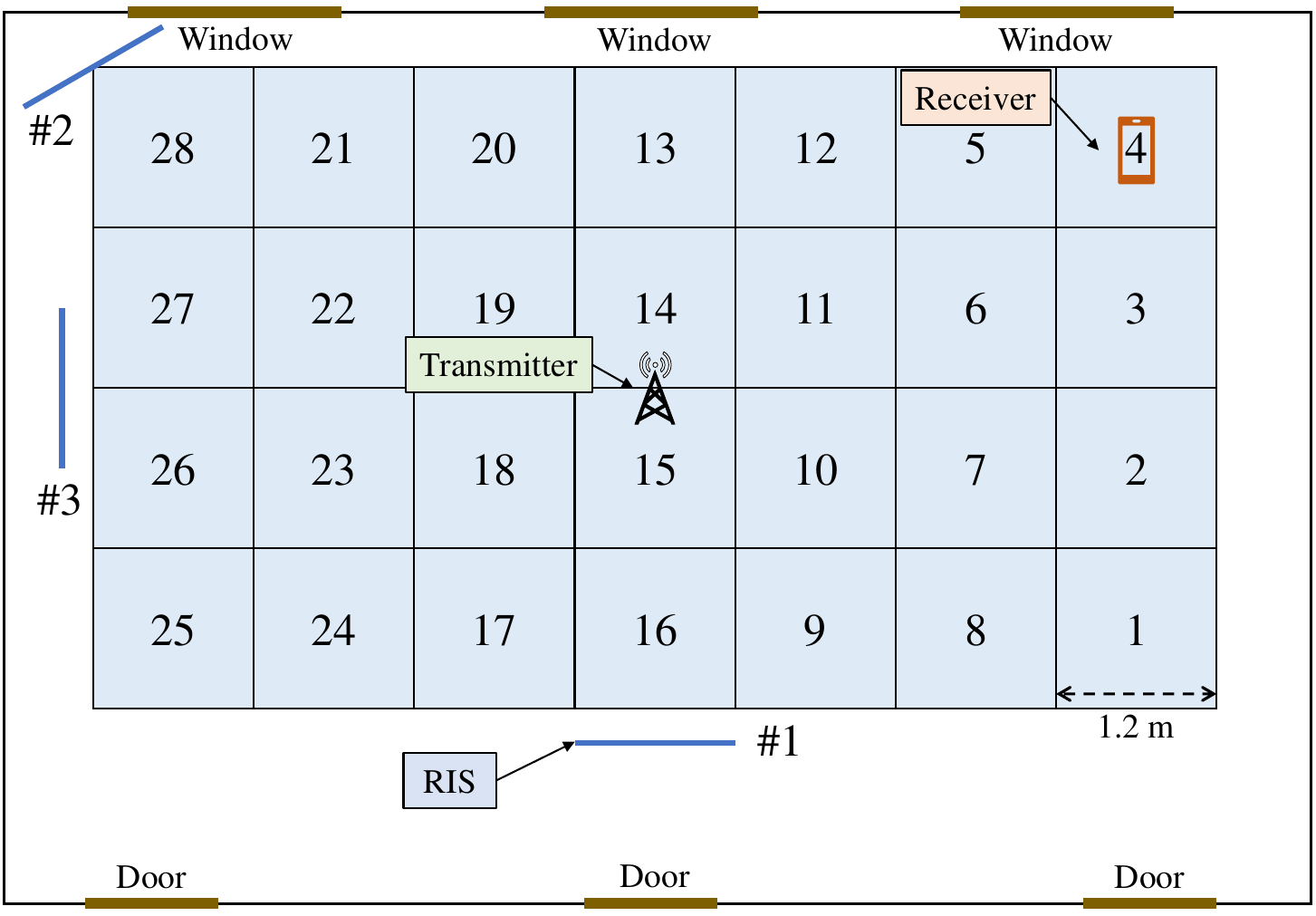}
		\caption{The floor plan of the office environment. The room is divided into $ 4 \times 7 $ squares. As marked in the figure, each square has a side length of 1.2 meters.}
		\label{fig:officeroom_map}
	\end{figure}
	
	\begin{figure}[t!]
		\centering
		\includegraphics[width=\linewidth]{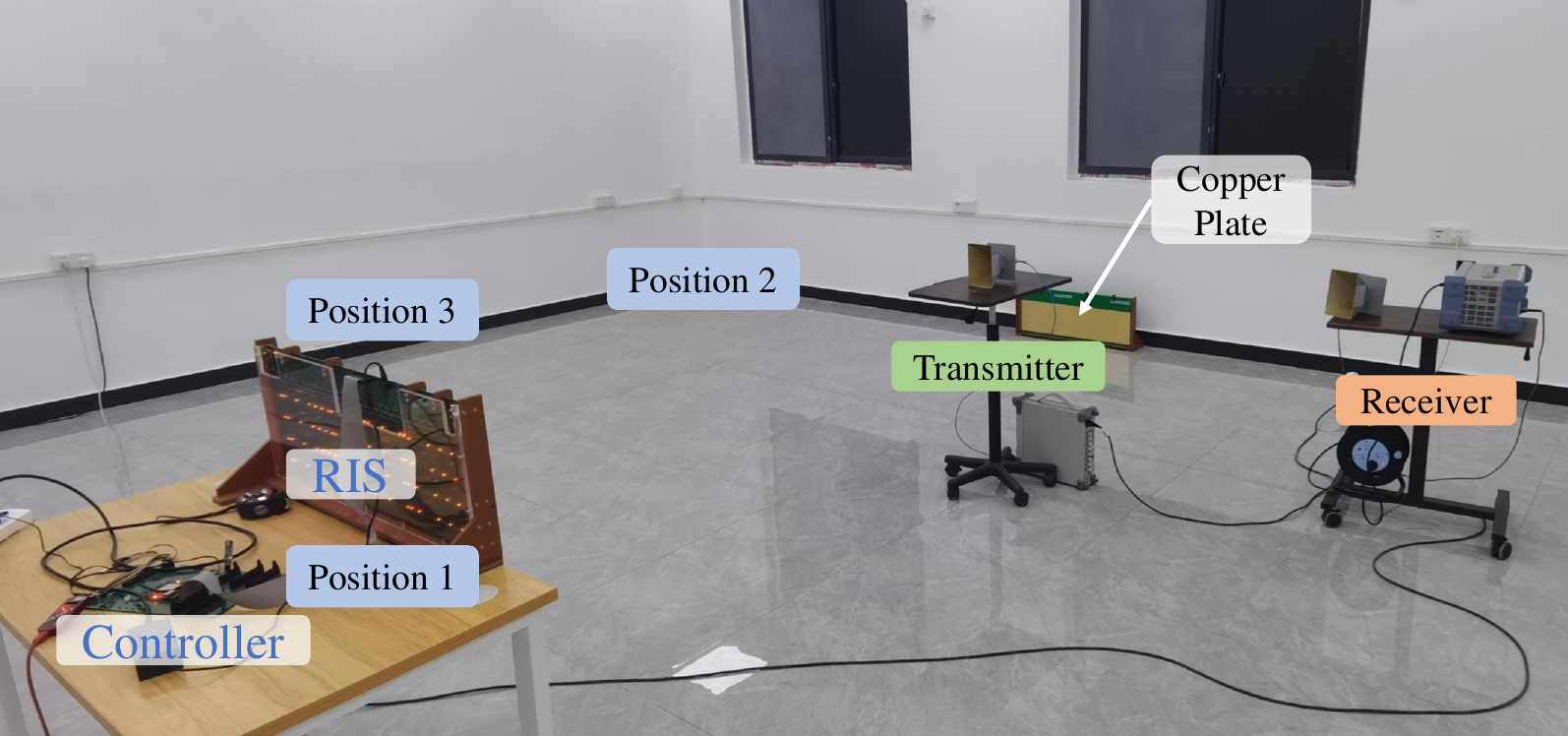}
		\caption{Measurement environment for the office scenario. The RIS is placed alternately in positions 1, 2, and 3.}
		\label{fig:officeroom_photo}
	\end{figure}
	
	\begin{figure*}[t!]
		\centering
		\includegraphics[width=\linewidth]{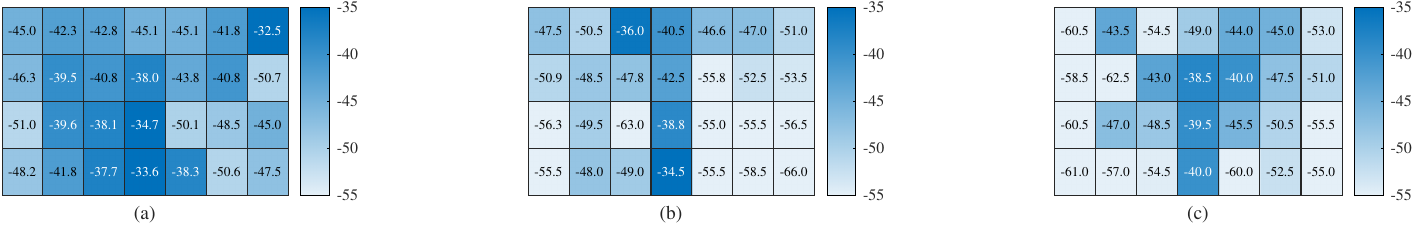}
		\caption{Test results for the office scenario at position 1: (a) point 4 enhanced by the RIS; (b) the RIS replaced by a copper plate; (c) nothing at position 1.}
		\label{fig:position1}
	\end{figure*}
	
	\begin{figure*}[t!]
		\centering
		\includegraphics[width=\linewidth]{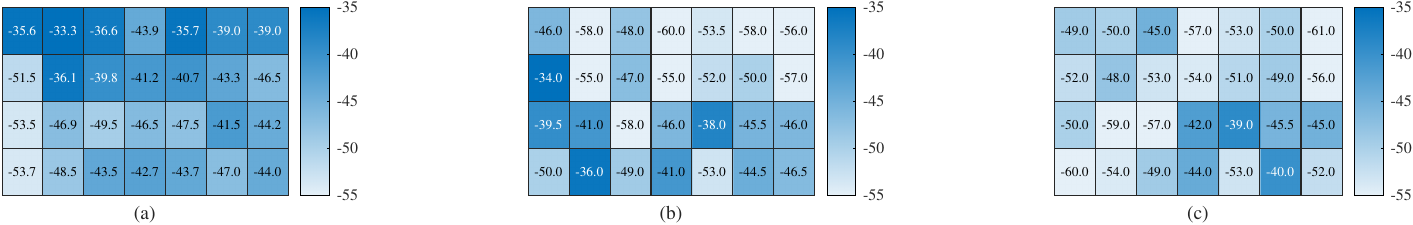}
		\caption{Test results for the office scenario at position 2: (a) point 4 enhanced by the RIS; (b) the RIS replaced by a copper plate; (c) nothing at position 2.}
		\label{fig:position2}
	\end{figure*}
	
	\begin{figure*}[t!]
		\centering
		\includegraphics[width=\linewidth]{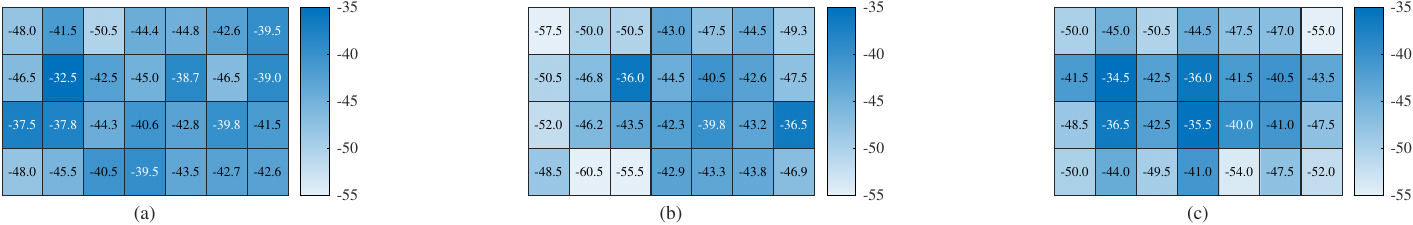}
		\caption{Test results for the office scenario at position 3: (a) point 4 enhanced by the RIS; (b) the RIS replaced by a copper plate; (c) nothing at position 3.}
		\label{fig:position3}
	\end{figure*}
	
	This experiment was conducted in an office room.
	The floor plan of the office environment is shown in Fig. \ref{fig:officeroom_map}.
	The room was divided into a grid consisting of $4 \times 7$ squares, with each square having a side length of 1.2 m.
	We placed the transmitter at the center of the grid area with a transmit power of +10 dBm.
	The RIS was positioned at three locations (1, 2, and 3) in the figure, and the horn antenna was directed from the transmitter towards the center of the RIS.
	Next, we configured the RIS to direct the reflected beam at point 4 and kept it unchanged throughout the experiment.
	Then, we moved the receiver along points 1 to 28 sequentially, kept it fixed for 30 seconds and recorded the received signal power.
	The measurement was then repeated with the RIS board replaced by a smooth metal plate of the same size as the RIS, as well as without any RIS or metal plate.
	
	\subsubsection{Position 1}
	We placed the RIS at the center of the long side (position 1) and measured the signal power at different points in the area.
	Fig. \ref{fig:position1} depicts the result of this configuration.
	The RIS provided a 20.5 dB gain over the baseline scenario (no RIS or metal plate) and an 18.5 dB gain over the copper plate scenario.
	The RIS exhibited a universal effect on signal enhancement, significantly boosting the signal power at the target point 4 as well as at many other points by varying degrees.
	Unlike the other two scenarios, there were no weak signal points in the entire area.
	
	\subsubsection{Position 2}
	The RIS was located at the upper left corner of the room.
	Fig. \ref{fig:position2} shows the experimental results.
	The RIS increased the received signal power by 22 dB compared to the scenario without the RIS.
	Due to the corner placement of the RIS, the received power of each point along the line between the RIS and point 4 was significantly enhanced.
	However, at points 25 and 26, the signal power was similar to that without the RIS and 15 dB lower than that with the copper plate.
	This was because these two points and the transmitting antenna formed a specular reflection through the copper plate, which was not at a 45-degree angle in the corner (see Fig. \ref{fig:officeroom_photo}).
	Moreover, these two points might be in the lobe depression of the pattern in the RIS-enhanced scenario.
	As a result, the receiving power would be quite low.
	Nevertheless, this does not imply that the RIS cannot optimize these poor signal points.
	To prove this, we reran the algorithm at point 25 to direct the beam to this location.
	The received signal power increased to -40.3 dBm, which was about 13 dB higher than before.
	
	\subsubsection{Position 3}
	The RIS was located in the middle of the short side of the room.
	The received power of each point is shown in Fig. \ref{fig:position3}.
	The RIS enhanced the received signal power by 15.5 dB compared to the scenario without the RIS, which was about 5 dB lower than in the previous two cases.
	The performance of the RIS in this position is poor due to the following two reasons:
	Firstly, the RIS-RX distance was the longest in the three scenarios when the receiver was at point 4.
	This caused the receiver to pick up more reflected signals from the environment, such as walls.
	At the same time, the energy received by the receiver from the RIS decreased (distance increased), which reduced the regulation ability of the RIS.
	Secondly, the reflection angle in this position was the smallest of the three.
	As a result, this position was closer to the specular reflection for the copper plate.
	Therefore, the enhancement effect of RIS deteriorated.
	
	Based on the findings of the above experiments, the RIS can achieve a 10 to 20 dB gain in an office environment under static conditions.
	Furthermore, the gains will increase further if RIS supports dynamic beam configuration.
	These results verify the signal enhancement capability of RIS in office settings.
	
	\subsection{Corridor environment}
	
	\begin{figure}[t!]
		\centering
		\includegraphics[width=\linewidth]{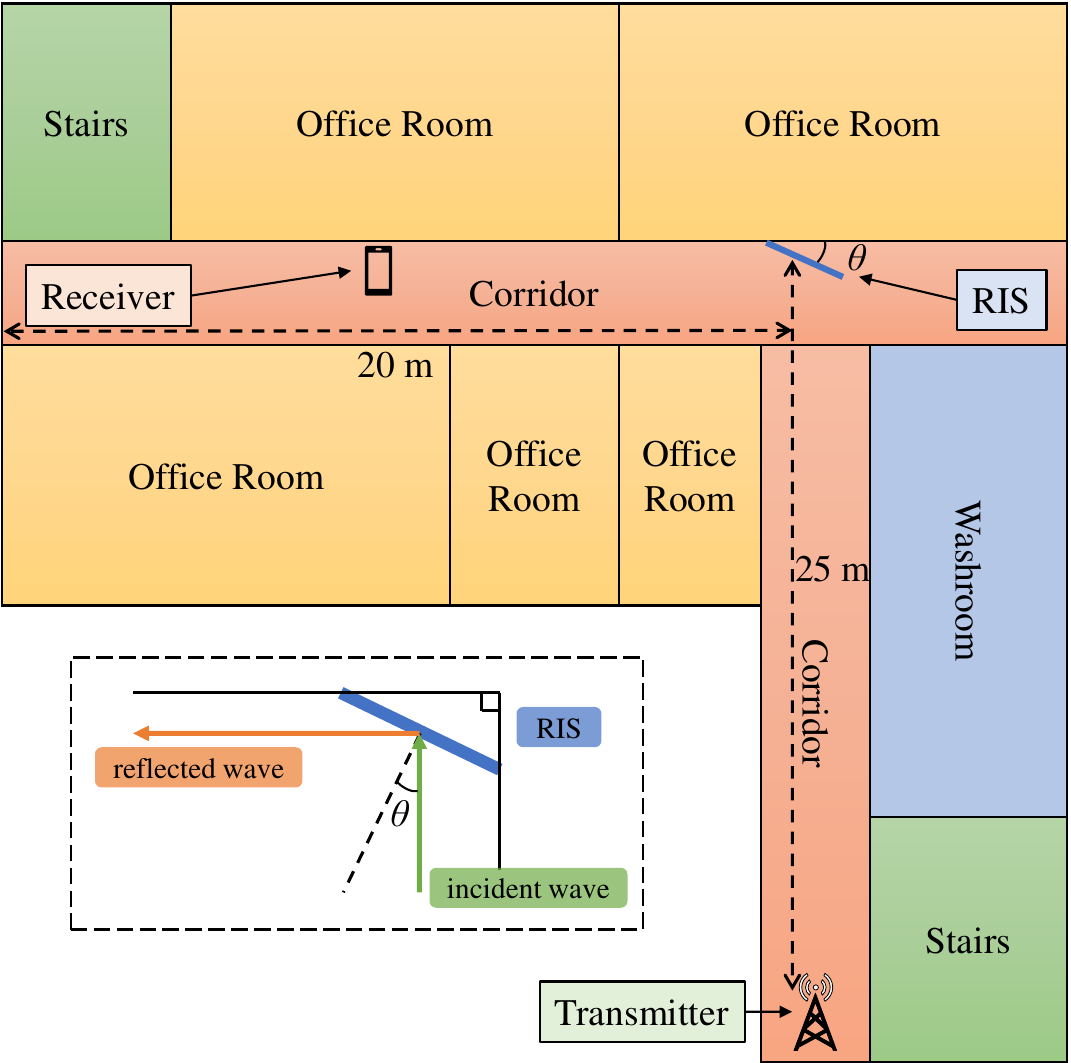}
		\caption{The floor plan of the corridor environment. The RIS is placed at the corner of the corridor.}
		\label{fig:corridor_map}
	\end{figure}
	
	\begin{figure}[t!]
		\centering
		\includegraphics[width=\linewidth]{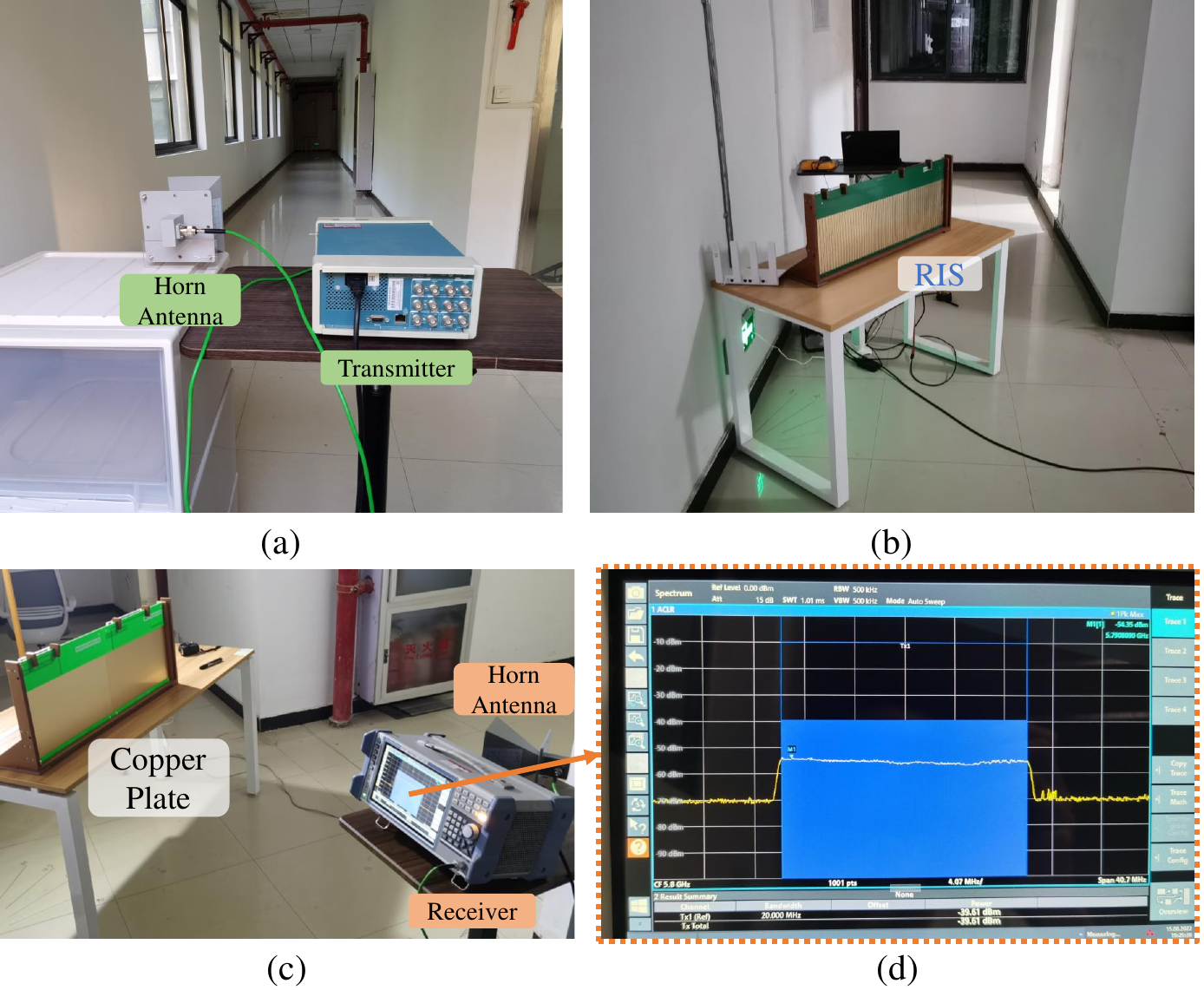}
		\caption{Photograph of the experimental setup: (a) the transmitter; (b) the RIS and the RIS controller (behind the RIS board); (c) the copper plate with the same size of the RIS and the receiver; (d) the view of the frequency spectrum analyzer.}
		\label{fig:corridor_photo}
	\end{figure}
	
	This experiment was conducted in an L-shaped corridor. Fig. \ref{fig:corridor_map} shows the floor plan of the corridor and its surrounding environments.
	The RIS was placed at the corner of the corridor, and the transmitter and receiver were located in the middle of the corridor. The angle between the RIS and the wall was $\theta$, which was also the incident angle. The width of the corridor is 2.11 m. The TX-RIS distance was 25 m, and the RIS-RX distance ranged from 0 m to 20 m with a step of 1 m. The transmit power was 0 dBm.
	
	Fig. \ref{fig:corridor_photo} shows the layouts of the corridor experiment, a typical hallway scene with windows, iron doors, and iron pipes in the ceiling. These reflectors create a complex environment for the corridor. A signal with a center frequency of 5.8 GHz and a bandwidth of 20 MHz was generated by the RF source. The spectrum analyzer measured the total received signal power throughout the entire frequency band.
	
\begin{figure}[t!]
		\centering
		\includegraphics[width=\linewidth]{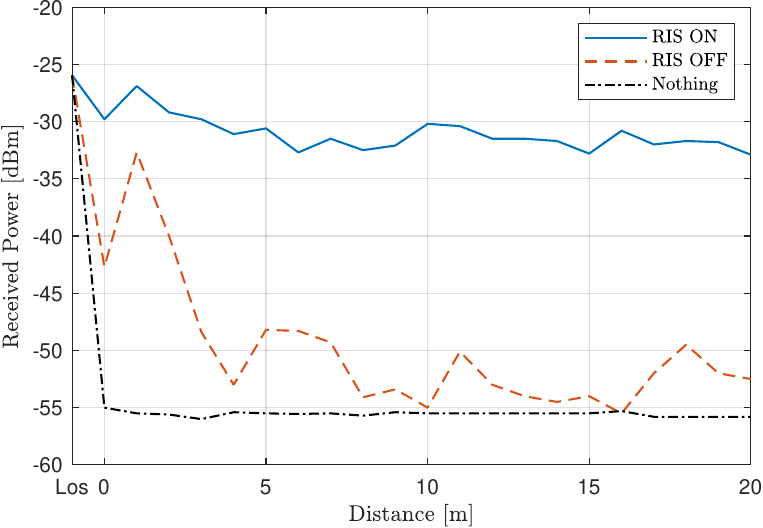}
		\caption{The received power vs. the distance between the receiver and RIS, $\theta=30^\circ$.}
		\label{fig:corridor}
	\end{figure}

	After setting up the test environment, we first placed the receiver at the end of the corridor. Then we measured the received power at different angles ($15 ^\circ, 30 ^\circ, 45 ^\circ, 60 ^\circ$ and $75 ^\circ$) with respect to the transmitter. The average results show that the RIS-on mode achieves a 14 dB gain compared to the RIS-off mode when $\theta$ is $15^\circ$ or $75^\circ$, a 20 dB gain when $\theta$ is $30^\circ$ or $60^\circ$, and a 6 dB gain when $\theta$ is $45^\circ$. Next we fixed $\theta$ at 30 degrees and ran the beamforming algorithm at the far end. Fig. \ref{fig:corridor} shows the test results. We observe that without the RIS, the received power drops by 30 dB compared to the line-of-sight (LoS) path.
	When the RIS is off, the received power fluctuates below -50 dBm as the distance increases. However, when the RIS is on, the received power remains around -32 dBm.
	It indicates that RIS can significantly enhance the received power in the corridor without a direct path, with a power increase of 6-20 dB, thus improving the communication quality.
	
	\begin{figure}[t!]
		\centering
		\includegraphics[width=\linewidth]{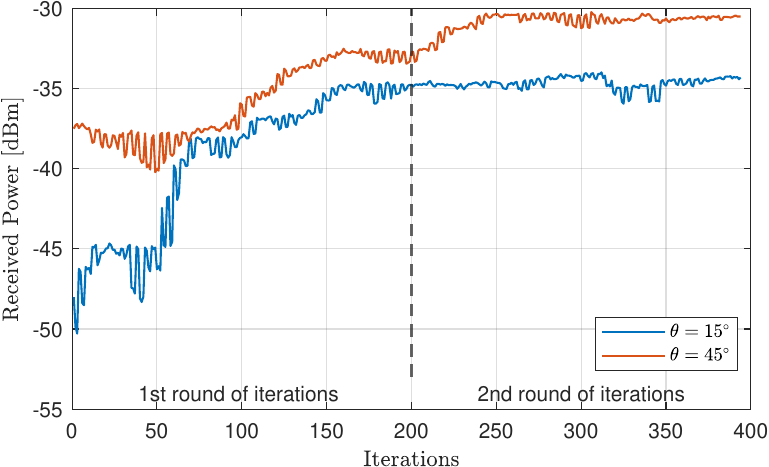}
		\caption{The received power as a function of time for the greedy fast beamforming algorithm (two iterations).}
		\label{fig:iteration}
	\end{figure}
	
	In addition, we recorded the temporal variation of the received power during the greedy fast beamforming algorithm \cite{pei2021ris}, as shown in Fig. \ref{fig:iteration}.
	The results show that the received power was close to the optimal value at the end of the first iteration, and reached and maintained the highest value in the second iteration.
	This suggests that the greedy fast beamforming algorithm can quickly converge.
	
	\subsection{Outdoor environment}
	
	\begin{figure}[t!]
		\centering
		\includegraphics[width=\linewidth]{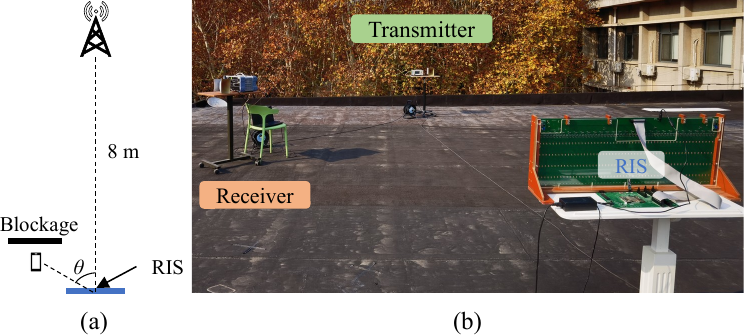}
		\caption{Outdoor test environment: (a) the floor plan consisting of the transmitter, receiver, RIS, and obstacle; (b) the photograph of the test.}
		\label{fig:outdoor_map}
	\end{figure}
	
	\begin{figure}[t!]
		\centering
		\includegraphics[width=\linewidth]{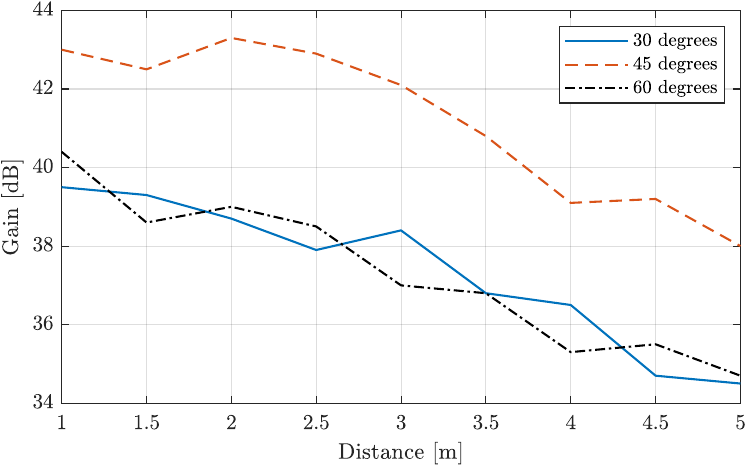}
		\caption{Test results of the outdoor environment.}
		\label{fig:outdoor_data}
	\end{figure}
	
	We also conducted an outdoor over-the-air test of the 5.8 GHz RIS system on the roof of our lab. Fig. \ref{fig:outdoor_map} (a) shows the floor plan of the environment, where the transmitter is 8 meters away from the RIS.
	The impinging signal is perpendicular to the surface in the experiment, and the angle of reflection is $\theta$. The photograph taken during the experiment is displayed in Fig. \ref{fig:outdoor_map} (b).
	We used a wall made of absorbing materials to simulate the occlusion of obstacles (according to individual experiments, the absorbent material wall can result in a 15 dB signal attenuation).
	Since there are fewer outdoor reflectors on this roof than in the previous scenario, we increased the transmit signal level to +16 dBm to observe the received signals when the RIS is absent.
	
	In the specific test, we measured the gain brought by RIS at various receiving distances (from 1 m to 5 m with 0.5 m intervals) and reflection angles ($ \theta = $ 30 degrees, 45 degrees, and 60 degrees) compared to the case without the RIS.
	Fig. \ref{fig:outdoor_data} displays the relationship between the gain brought by RIS and the RIS-RX distance.
	The gain for all three angles decreases as the distance increases.
	On average, the gain is 41.2 dB at a reflection angle of $45^\circ$, and 37.3 dB at $30^\circ$ and $60^\circ$.
	It can be seen that under the same RIS-RX distance, the gain with a reflection angle of $45^\circ$ is higher than $30^\circ$ or $60^\circ$. 
	
	There are several potential causes for this observation. 
	First, when the reflection angle is less than 45 degrees, the TX, RIS and RX are close to the mirror reflection.
	Nearby objects, such as the RIS bracket and the table supporting the RIS, provide alternative reflection paths for the signal, resulting in high received signal power without using the RIS.
	When the reflection angle exceeds 45 degrees, the performance of the RIS decreases, attributed to the reduction in effective surface area of the RIS.
	Therefore, the optimal gain effect can be achieved around 45 degrees.
	In addition, as the distance increases, the field of view of RIS decreases at the receiver, and the scattering signal from the environment increases.
	These factors reduce the gain provided by the RIS.
	Nevertheless, it is remarkable that the RIS can offer more than 35 dB gain at these angles and distances.
	The experiment demonstrates the potential of the RIS in coverage enhancement of wireless signals.
	
	\section{2.6 GHz RIS Field Trials}\label{sec:2.6GHz_RIS_test}
	In this section, we present the results of our experiments with our 2.6 GHz RIS in a laboratory setting and in a commercial 5G network environment.
	
	\subsection{Amplitude and Phase Frequency Response Test}
	
	\begin{figure}[t!]
		\centering
		\includegraphics[width=\linewidth]{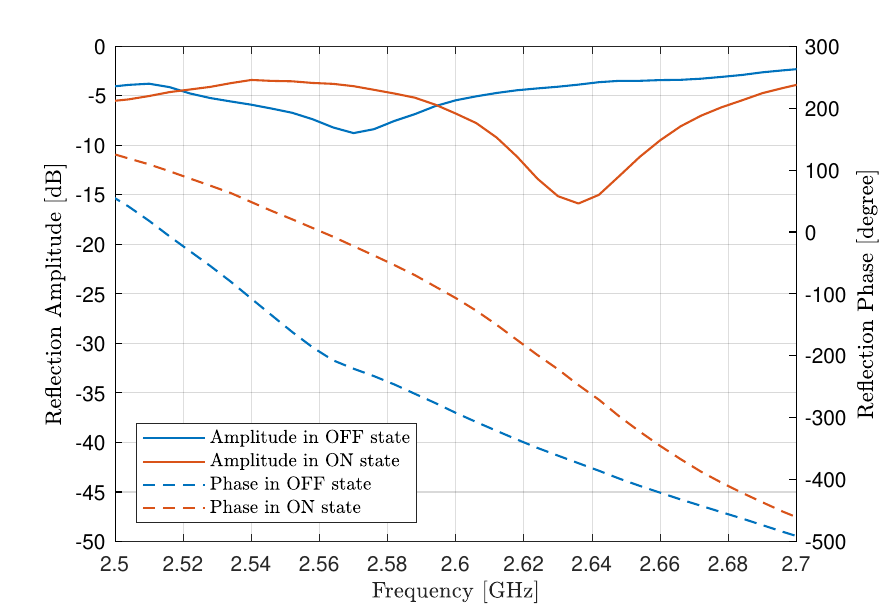}
		\caption{Measurement result of the unit cell in OFF/ON states. The bias voltage of the diode is 0 V in the OFF state and 5.5 V in the ON state. When the frequency is between 2.55 GHz and 2.61 GHz, the phase shift of the two states is between 160 degrees and 200 degrees.}
		\label{fig:2.6GHz_frequency_phase}
	\end{figure}
	
	The electromagnetic characteristics of a specific RIS may differ from the simulation results due to the factory manufacturing process.
	Consequently, it is necessary to measure the characteristics of the RIS and select the suitable operating voltage based on the obtained test results.
	The amplitude-frequency and phase-frequency characteristics can be conveniently measured using a VNA.
	During the testing phase, both the transmitting and receiving antennas were positioned at a distance of 1 meter from the RIS.
	Following the acquisition of the S12 parameter, the target voltages of 0 V and 5.5 V were determined by comparing the phase difference.
	
	As shown in Fig. \ref{fig:2.6GHz_frequency_phase}, the designed RIS unit and the control circuit meet the design requirements under the selected voltages.
	The RIS unit can produce a phase shift of about 200 degrees at 2.6 GHz for two voltage states. 
	Moreover, in the frequency range of 2.55 GHz to 2.61 GHz, the phase difference is between 160 degrees and 200 degrees, and the amplitude curve is relatively flat.
	This capability is a prerequisite for enabling signal enhancement in commercial networks on the China Mobile N41 5G band (2515-2675 MHz).
	
	\subsection{Near-Field Trials}
	
	\begin{figure}[t!] 
		\centering
		\includegraphics[width=\linewidth]{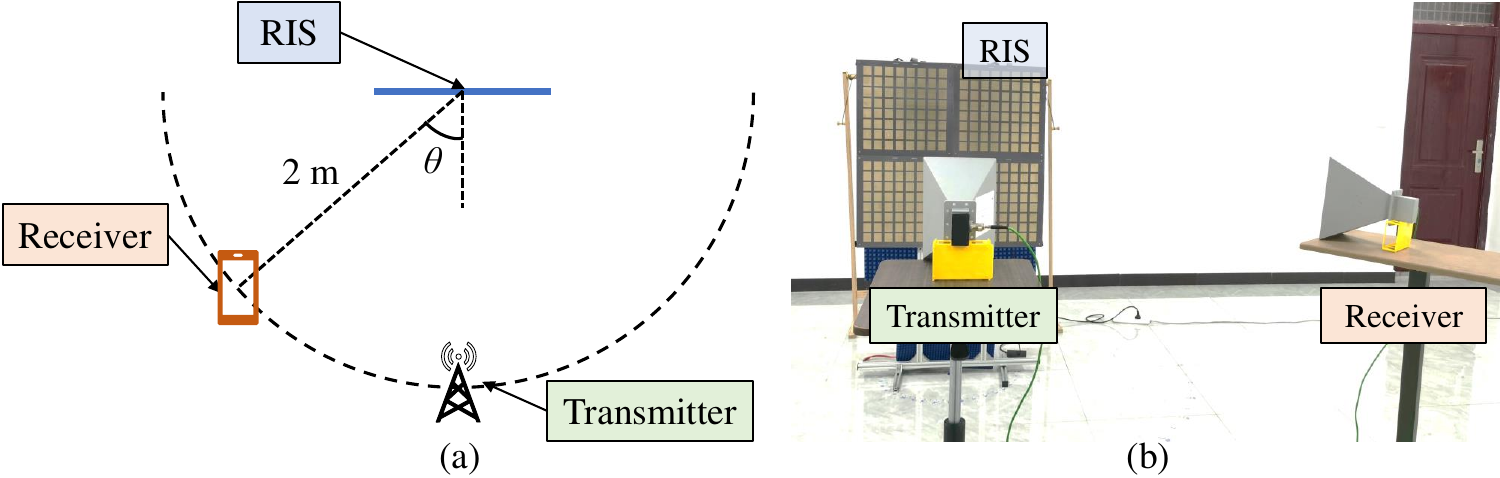}
		\caption{Near-field test environment: (a) the positions of the RIS, transmitter and receiver; (b) the photograph of the test.}
		\label{fig:near}
	\end{figure}
	
	We performed a near-field test with the 2.6 GHz RIS in a laboratory setting.
	Fig. \ref{fig:near} illustrates the experimental setup.
	The RIS was positioned at the center of the room, facing the transmitting antenna that was 2 meters away from the RIS panel.
	The receiving antenna moved along a semicircle with a radius of 2 meters from the RIS center, and the angle between the receiving antenna and the RIS normal ranged from 30 degrees to 60 degrees.
	This was a typical near-field scenario as the side length of the $2 \times 2$ RIS composite plate has reached 90 cm.
	We used the DFT codebook, the angle-based two-dimensional codebook and the two-step spatial oversampling codebook for beamforming.
	
	Table \ref{tab:2.6GHz_Near} shows the average performance of these methods.
	The angle-based two-dimensional codebook achieves about 1 dB higher gain than the DFT codebook.
	Moreover, the proposed two-step spatial oversampling codebook performs similarly to the previous algorithm, yet with much lower complexity,  which validates its effectiveness.
	
	\begin{table}[t!]
		\renewcommand{\arraystretch}{1.3}
		\caption{Power Gains under Different Algorithms}
		\label{tab:2.6GHz_Near}
		\centering
		\begin{tabular}{|c|c|}
			\hline\hline
			                 \textbf{Method}                 & \textbf{Average Gain} \\ \hline
			                  DFT codebook                  &        10.5 dB        \\ \hline
			              Angle-based two-dimensional codebook              &        11.7 dB        \\ \hline
			     Two-step spatial oversampling codebook      &        11.5 dB        \\ \hline\hline
		\end{tabular}
	\end{table}
	
	\subsection{Coverage Enhancement Test in Commercial 5G Networks}
	
	\begin{figure}[t!]
		\centering
		\includegraphics[width=\linewidth]{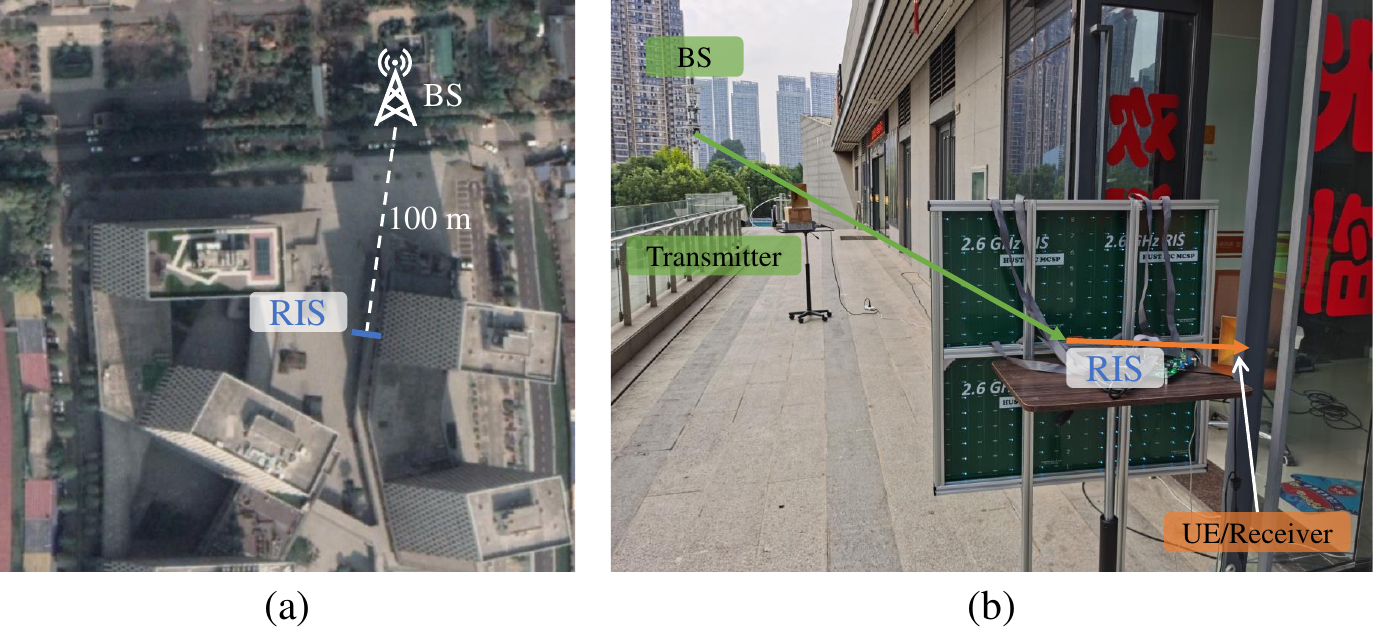}
		\caption{Photographs of the test environment: (a) the satellite map; (b) the experimental setup, including the BS, transmitter, RIS, UE, and receiver.}
		\label{fig:outdoor_photo_2.6GHz}
	\end{figure}
	
	\begin{figure}[t!]
		\centering
		\includegraphics[width=0.6\linewidth]{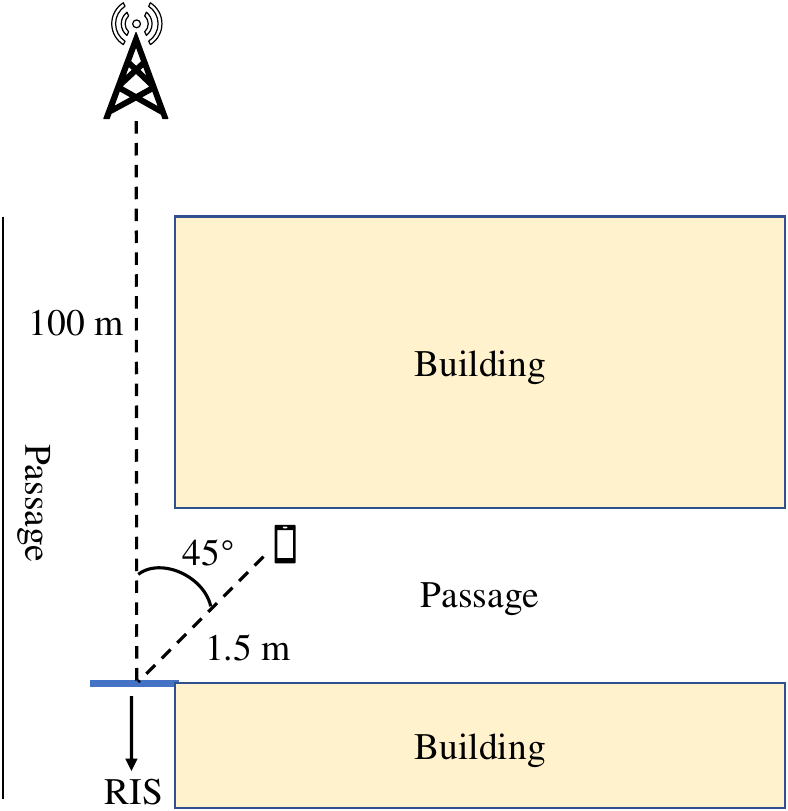}
		\caption{The top view of the test environment. The RIS is installed on the the walkway outside the building with a LoS path to the BS, and the receiving phone is located on the sidewalk inside the building.}
		\label{fig:outdoor_map_2.6GHz}
	\end{figure}
	
	To evaluate the performance of the 2.6 GHz RIS system under the commercial network, we conducted an experiment in a large industrial park.
	Fig. \ref{fig:outdoor_photo_2.6GHz} shows the photographs of the test environment.
	The BS was located 100 meters away from the passage entrance.
	Fig. \ref{fig:outdoor_map_2.6GHz} illustrates the positions and orientations of the BS, RIS, receiving phone, and buildings.
	The phone and RIS were 1.5 meters apart and formed a 45-degree angle in the normal direction.
	During the experiment, we used a signal-testing software to monitor the signal reception from the BS and record the strength.
	
	We first analyzed the cell list on the software without deploying RIS and found that the BS cell was not in the list when the UE was inside.
	Then, we placed the RIS at the entrance of the passage with the antenna side facing the BS.
	The UE could decode the physical cell ID (PCI) of the target cell. However its Reference Signal Receiving Power (RSRP) was lower than -90 dBm.
	We observed that the RSRP fluctuated by approximately 2 dB over time, possibly due to the interference of environmental factors such as trees and wind.
	This variation adversely affected the performance of the greedy fast beamforming algorithm.
	
	The angle-based codebook approach proposed in Section \ref{sec:prototypes} is able to alleviate this problem.
	It enables us to determine the codeword based on the reflection angle and the corresponding steering vector.
	We can measure these angles and provide feedback to the controller.
	Alternatively, We may scan the codebook with a predefined series of angles sequentially to find the codeword with the highest RSRP.
	Following multiple iterations of testing and averaging, we discovered that indoor signal strength has increased by 4 to 7 dB when the RIS was powered on compared to when it was powered off.
	The experimental results were approximately 10 dB worse than the previous near-field experiment.
	To identify the causes of this discrepancy, we conducted a comparative experiment. We used an RF signal source and a spectrum analyzer as the transmitter and receiver, respectively.
	The transmitter was located 10 meters away from the RIS, along the line connecting the BS and the RIS center, while the receiver was at the original position of the mobile user.
	We performed the experiment with both omnidirectional and directional antennas.
	The directional antenna yielded a similar gain to the near-field test, around 15 dB. However, the omnidirectional antenna only provided a 4 dB gain.
	This result is consistent with the actual BS-to-UE scenario, and it poses a challenge to enhance the gain in practical omnidirectional antenna environments.
	The preliminary test results suggest that the RIS can adapt the signal beam according to the position of the user and improve signal strength and communication quality in the poorly served areas of commercial network.
	
	\section{Conclusion}\label{sec:conclusion}
	In this paper, we designed and fabricated a prototype of RIS-aided wireless communication system operating at 2.6 GHz band.
	We also proposed a low-complexity codebook algorithm based on the spatial structure of the channel.
	Furthermore, experimental results demonstrate that the two-step spatial oversampling codebook greatly reduces the searching time while having a negligible performance loss compared to the 2-D full-space searching.
	We evaluated and tested the RIS systems operating at 5.8 GHz and 2.6 GHz in multiple scenarios.
	In the office and corridor scenarios, the 5.8 GHz RIS provided a power gain of 10-20 dB at the receiver, while in the outdoor test, it achieved a power gain of 35 dB compared to non-deployment case.
	The 2.6 GHz RIS improved indoor coverage by 4-7 dB in commercial 5G networks.
	These experimental results reveal that RIS achieves higher power gain if the transceivers are equipped with directional antennas instead of omni-directional antennas. 
	
 
	\bibliographystyle{IEEEtran}
	\bibliography{references}
	%
	%
\end{document}